\documentclass[preprint,12pt]{elsarticle}
\usepackage{amsmath,amssymb,amsfonts}
\usepackage{algorithm}
\usepackage{algcompatible}
\usepackage{graphicx}
\usepackage{textcomp}
\usepackage{xcolor, soul}
\sethlcolor{yellow}
\usepackage{hyperref}
\usepackage{tikz}
\usepackage{stfloats}
\usepackage{placeins}
\usepackage{tabularx}
\usepackage{tabularray}
\usepackage{float}
\usepackage{tablefootnote}
\usepackage{afterpage}
\usepackage{geometry}

\usepackage[labelfont=bf,justification=raggedright,singlelinecheck=false]{caption}
\usepackage{amssymb}

\begin{document}

\begin{frontmatter}



\title{HexE - Securing Audio Contents in Voice Chat using Puzzle and Timestamp}


\author[inst1]{Aadhitya A}

\affiliation[inst1]{organization={Department of Information Technology},
            addressline={Anna University}, 
            city={Chennai}, 
            country={India}}

\begin{abstract}
Cryptography is the study of securing information. It is the physical process that scrambles the information by rearrangement and substitution of content, so that it becomes difficult for anyone to understand. In today's world, security has become an inevitable part of our day-to-day life, right from normal browsing to performing critical payment transactions. Hackers work endlessly to break the security present in the apps/websites on which we perform day-to-day operations and salvage valuable information. Because of this, many illegal activities have taken place which affect the user. One such illegal activity is tapping the voice communication between two users. If left unencrypted, the communication between the users is compromised, thereby causing issues. One way to prevent this act is to encrypt the audio in that the contents cannot have tampered with unless the receiver has the valid key to decrypt it. The proposed solution termed "HexE" aims to create a puzzle-based algorithm which would encrypt and decrypt the audio files without manipulating the file header, thus securing the contents. The algorithm works on an NxN SuDoKu-based puzzle which is accepted both by the sender and receiver. Using the timestamp of the event (UNIX based), a grid from the puzzle is chosen which in turn will act as the key for both encryption and decryption. If the timestamp is slightly adjusted, the process will end up in failure during decryption, thus ensuring confidentiality. Another approach to secure the audio files is to implement IPFS (Inter Planetary File System) alongside the puzzle algorithm in which the encrypted audio is stored on it and the receiver can fetch the audio provided if the valid IPFS Hash of the file is present. In this way, the audio file is secured.
\end{abstract}




\end{frontmatter}


\section{Introduction}
In today’s world, there is a growing need for secure communication in various industries, including government, military, and financial sectors. One way to secure communication is using encryption, which is the process of converting plaintext (i.e., unencrypted data) into ciphertext (i.e., encrypted data) that can only be decrypted with the correct key. While there are many techniques for encrypting and decrypting data, a new method can be used that uses Sudoku puzzles as a key for encrypting and decrypting audio files in the WAV format.

There are several research challenges related to making an audio encryption and decryption algorithm:
Developing a robust and secure algorithm: One of the main challenges in audio encryption and decryption is developing algorithms that are both effective and secure. This involves finding ways to protect the original audio data while also making it difficult to decrypt without the correct key.

Ensuring high security: Another important research challenge is ensuring that audio files are secure and resistant to attack. This may involve designing algorithms that are resistant to common cryptographic attacks, such as brute-force attacks and known-plaintext attacks.

Improving efficiency: Another potential research challenge is finding ways to improve the efficiency of the audio encryption and decryption process. This may involve optimizing algorithms for faster performance, or finding ways to reduce the computational resources required to perform the encryption and decryption.

Integration with other systems: Finally, one potential research challenge is integrating audio encryption and decryption systems with other systems, such as audio editing software or communication platforms. This will involve finding ways to seamlessly integrate the system with these other systems, while also maintaining the security and effectiveness of the encryption and decryption process.

Sudoku is a popular logic puzzle that involves filling in a grid of squares with numbers so that each row, column, and region contains all the numbers from 1 to N. While Sudoku can be a fun and engaging activity, it is not typically in use for generating unique binary keys for encryption. There are many different methods for generating unique binary keys from it. One such method is taking the mini-grid of the puzzle based on the values obtained from the operations performed on the UNIX timestamp. After modification of the mini-grid with the values obtained, the binary key is generated which is then XORed with audio contents for the encryption and decryption process.

The goal of the proposed project is to develop an algorithm that can perform encryption and decryption on WAV audio files using Sudoku puzzles as the key. The system should be able to take a WAV audio file and a Sudoku puzzle as input, and output an encrypted version of the audio file. It should also be able to take the encrypted audio file and the original Sudoku puzzle as input, and output the decrypted version of the audio file. The main challenges include developing a robust and secure encryption algorithm, ensuring high security, improving efficiency, and making the system user-friendly and easy to integrate with other systems. The final system should be able to overcome the mentioned challenges and perform encryption and decryption tasks efficiently and accurately, with a high level of security.

The section-wise organisation of the paper is as follows: Section 1 gives a brief introduction about creating a proposed algorithm called "HexE" for audio encryption and decryption using NxN Sudoku puzzle and timestamp. Section 2 analyses and gets various notions from international conference and journal papers. Section 3 describes modules to be implemented, the system architecture, system design of the proposed work and also describes the execution details of the modules proposed. Section 4 presents the evaluation results tested under different scenarios and different files. Section 5 combines the conclusion of the proposed work and future enhancements that can be deployed in proposed work.

\section{Related Work}
Hashim et al., (2018) \cite{b1} proposed a steganography method in which the LSB bits of the audio are extracted and encrypted using the AES algorithm. The stego audio obtained was similar to the original audio but had limitations based on the key used during the process. Phipps et al., (2022) \cite{b2} experimented with various voice samples in which the message can be hidden in the audio files and can be sent in the audio transmission channels which would enhance the biometric authentications. Ying et al., (2021) \cite{b3} proposed an adaptive Syndrome-Trellis-Code method in which the hidden messages can be embedded into the audio files with minimal distortion in the audio quality. Ahmed et al., (2020) \cite{b4} made a review on digital steganography on audio files and listed out methods in which steganography can be done on audio, its merits, demerits and so on. Mahmoud et al., (2022) \cite{b5} proposed an LSB-BMSE method for audio steganography in which the message is encoded first and then the BMSE method is applied to the LSB of audio to hide the message.

Ashari (2021) \cite{b6} made research in which the image can be hidden in the cover audio and extracted the image back without much loss of content. Dudhwal et al., (2021) \cite{b7} made a survey on audio steganography based on the LSB method and described the advantages, and disadvantages of the method. Alhassan et al., (2022) \cite{b8} proposed a method in which the message can be hidden in audio files using a novel twin K-Shuffling technique. The message was able to hide the audio but the cover audio used must be of a certain length to hide the message efficiently. Ignacio et al., (2022) \cite{b9} made an experiment in which the fingerprint is extracted from the audio file and the saliency map obtained after processing is stored on a database for many purposes like biometrics, and authentication. Abduljaleel et al., (2021) \cite{b10} proposed an algorithm that includes hybrid transformation to analyze the speech signal frequencies. The speech signal is then compressed, after removing low and less intense frequencies, to produce a well-compressed speech signal and ensure the quality of the speech. The proposed algorithm proved to be highly efficient in the compression and encryption of the speech signal based on approved statistical measures. 

Shukla et al., (2018) \cite{b11} proposed an encryption method where the audio files are encrypted by using genetic algorithms on converted waveform images. The proposed method worked well and stood against simulated attacks very well. Nurdin et al., (2022) \cite{b12} proposed a cryptool-based twofish algorithm to secure voice chat apps. The app is tested on live IP addresses and the algorithm handled well without the loss of audio content. Hemanth et al., (2022) \cite{b13} used modified XOR technique for audio steganography. Using XOR, the message to be sent is hidden well by embedding it within the audio in contrast to the LSB technique. The algorithm handled well and results are good compared with other LSB techniques. Abomhara et al., (2022) \cite{b14} used AES Encryption method to handle the encryption and decryption of video files having H.264/AVC Codec. AES handled pretty well but the computational complexity of the encryption and decryption process was high. Roy et al., (2022) \cite{b15} used a combination of hyperchaotic masking and modulation in order to encrypt speech communication on devices. The algorithm handled well and the error margin was less.

Razmara et al., (2022) \cite{b16} presented the design of an analog time-varying audio cryptography system that is based on sliding mode synchronization of non-identical chaotic systems. The system is described with time-delayed fractional-order dynamics. The proposed system is designed to ensure secure communication of audio signals in real-time. The authors used chaotic systems with non-identical dynamics as the basis for the cryptography system, which provides a high level of security due to the complexity and sensitivity to initial conditions of chaotic systems. The results show that the system is able to transmit audio signals securely, even in the presence of noise and other disturbances. Palaniappan (2022) \cite{b17} presented a unique design of cryptographic algorithm which is specifically designed for Auditory cryptography and visual cryptography to make the encryption and decryption technique stronger. This algorithm is a combination of multiple techniques such as Ant Algorithm, Logical Gates Technique, Dual authorization PINs, Indexed Arrays. Combination of these techniques makes the algorithm unique and strong to secure the data. This research was implemented on audio files, images and video files. The study of the result shows effective way of masking the data as it is hard to decode without PINs. Also, performance of the algorithm is efficient during encryption and decryption process. Gadde et al., (2023) \cite{b18} proposed an audio cryptographic algorithm where a combination of symmetric and asymmetric methods are utilized. To assure the confidentiality and integrity of medical data, an improved Robust S-box-based Advanced Encryption Standard (IRS-AES) is proposed with Runge-Kutta Optimization (RKO) algorithm. The Deoxyribonucleic Acid (DNA)-based Modified Elliptic Curve Cryptography (MECC) algorithm is introduced for key generation, and the best key is selected with Bald Eagle Search optimization (BES) algorithm. Finally, the medical data are encrypted IRS-AES algorithm and stored in the cloud. The performance of the proposed approach is enhanced with the communication overhead of 11.51 for the image dataset. MSE and PSNR obtained are 16.2 and 8.465 for an audio dataset, 75.21 and 3.5 for the video dataset.

Boumaraf et al., (2022) \cite{b19} presented an encryption scheme for AMR-WB ITU-T G.722.2 speech based on ECC for securing transmitted speech signals. The implementation of ECC is carried out by transforming the coded speech into an affine point on the elliptic curve, over a finite field. ECC has been shown to offer an RSA- grade security with smaller key size. Results show that ECC has advantage in terms of simplicity and security but has the disadvantage of larger processed file size. Gera et al., \cite{b20} (2022) used a new higher recognition Least Significant Bit (LSB) to conceal the data in audio. This technique is used to embed the hidden audio into cover audio of the same size. The strategy outperforms similar studies by enhancing hiding capability up to 30\% and preserving stego audio transparency with the SNR value at 72.2 and SDG at 4.8. Hemanth et al., (2022) \cite{b21} proposed LSB+XOR method where XOR is introduced to overcome the disadvantages of LSB method in audio cryptography. It is observed that the security of the transmission is highly increased by using this method. 

Arpita et al., (2022) \cite{b22} used pseudo random noise signal to encrypt the audio signal. The generation of pseudo random numbers are done by two interconnected Linear Feedback Shift Registers, the first of which controls the second. Different formulae are presented depending on the seed word, and therefore appropriate random numbers are generated. The usage of two LFSR increases the randomness of the random number generated and also increases the periodicity of generated random numbers making it more secure.The encrypted histogram shows better signal distribution like white noise which ensures more security. Bonny et al., (2023) \cite{b23} presented a new, highly secure chaos-based secure communication system that combines a conventional cryptography algorithm with two levels of chaotic masking technique. To enhance the security level, the authors employed the characteristic of a unified hyper-chaotic system to generate three different types of attractors. The simulation and comparison results demonstrate the high efficiency of the suggested cryptosystem and robustness against various cryptographic attacks. Shahriyar et al., (2022) \cite{b24} proposed a novel Speech Security System (SSS) which leverages DNA cryptography to provide security to the transmitted speech signal. The proposed method encrypts the speech signal into a DNA strand which works as a camouflage to the original speech signal. The DNA strand is decrypted at the receiver end and the original speech signal is retrieved. The proposed SSS is significantly robust against any cryptanalytic, statistical, and brute force attacks by an intruder and thus improves the security of the speech signals. Chen et al., (2022) \cite{b25} investigated several audio-based security techniques to secure Device-to-Device (D2D) communications and proposed a new method for secure audio communication. A novel audio-based secure transmission scheme was proposed by leveraging secure polar code and self-jamming techniques to achieve high efficiency and high level of security. Open research issues of audio-based security are also discussed to enhance the security of D2D communications.

\section{Proposed Work}
In the age of modern life, security becomes a viable part in every application and in every sector. This applies to the scenarios where the data must be transmitted in secret. Thus, the complexity of the cryptographic algorithm must be kept in mind as well. To overcome this, the user is asked to provide the level of encryption/decryption to be performed (based on grid size of the puzzle). Then, the audio file is sent to the file along with a pre-generated puzzle grid and UNIX timestamp. Using the components, the bytes of the audio file are altered to make it inaudible to the 3rd party user. If the receiver decrypts using the right components, the audio file is restored without any loss of quality.

\subsection{Architecture Design}
\begin{figure}[htp!]
	\centerline{\includegraphics[scale=0.5]{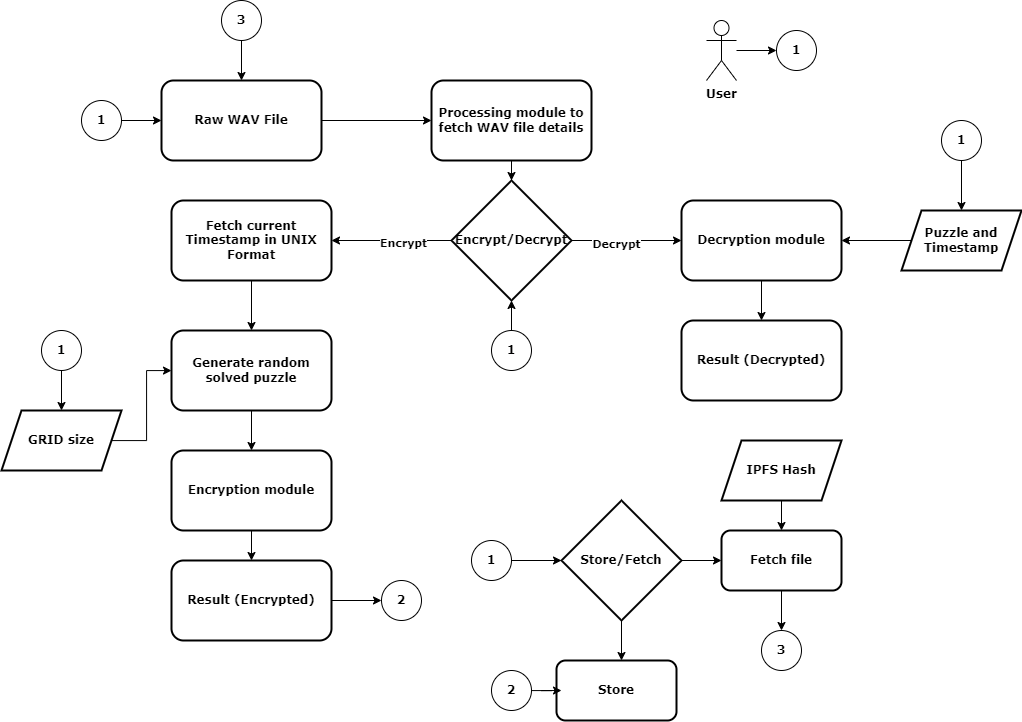}}
	\caption{Architecture for the proposed cryptographic system}
	\label{arch-0}
\end{figure}

The proposed system architecture will have 2 components: Algorithm and IPFS. The proposed algorithm will act as the core part of the system. The required components of the algorithm include a NxN sudoku puzzle, WAV audio file, UNIX timestamp. The IPFS module serves as an add-on for the algorithm and it is made optional to the end users for those who want to secure the audio file and store it securely. If the receiver knows the IPFS hash, then they can download it from the IPFS Gateway. 
The architecture diagram for the proposed algorithm is given in \ref{arch-0}

\FloatBarrier
\subsection{Tools Used}
The cryptographic algorithm was executed in using the following tools and frameworks:
\begin{itemize}
	\item Language
	\begin{itemize}
		\item Java
	\end{itemize}
	\item IDE
	\begin{itemize}
		\item IntelliJ IDEA
		\item Microsoft VS Code
	\end{itemize}
	\item Unit Testing
	\begin{itemize}
		\item JUnit 5
	\end{itemize}
	\item Audio Processing
	\begin{itemize}
		\item Java Sound API
	\end{itemize}
	\item Audio Analysis
	\begin{itemize}
		\item Sonic Visualizer
		\item MATLAB Online
	\end{itemize}
	\item IPFS
	\begin{itemize}
		\item Pinata IPFS
		\item Pinata IPFS Java API
		\item Cloudflare IPFS Gateway
	\end{itemize}
	\item Framework
	\begin{itemize}
		\item Maven
	\end{itemize}
	\item Hardware
	\begin{itemize}
		\item Intel Core i5 10th Gen
		\item 8GB DDR4 RAM
	\end{itemize}
	\item OS
	\begin{itemize}
		\item Windows 11
		\item Ubuntu 22.04
	\end{itemize}
\end{itemize}

\subsection{Algorithms}
The most important step in the system is to create an audio encryption algorithm. For any cryptographic algorithm, it consists of main two parts: Encryption and Decryption. The cryptographic algorithm must be designed in such a way that the algorithm must be efficient on time and memory resources. The procedural steps for the proposed algorithm are given below:
\begin{enumerate}
	\item First, a random solved Sudoku puzzle is generated and taken for generating a cipher key
	\item Then, the UNIX timestamp of the current time (during function execution) is taken, through which the variables p, k, t, u and u1 are calculated, where,
	\begin{itemize}
		\item p = Sum of digits present in the timestamp
		\item k = Square root of the size of puzzle
		\item t = p mod (size of sudoku puzzle)
		\item u = p mod k
		\item u1 = (Last digit of timestamp) mod k
	\end{itemize}
	\item Based on the above variables taken, the mini-grid is chosen for generating the cipher key, and the variables 'u' and 'u1' corresponds to the row and column of the 3x3 grid to be chosen
	\item The grid values are multiplied with variable 't' (if t=0, then it's taken as 1)
	\item The grid values are then converted to binary format and combined as a single binary string (in row-major order)
	\item Then, the audio file supplied by the user is converted to byte array format. The pointer of the loop is set to the 41st position as it denotes the data of the WAV audio file.
	\item Finally, the frames are XORed with the generated binary cipher key. In the end, the bytes are then converted to WAV audio file which will be playable but with contents distorted
	\item For decryption, the process is simply done again but the Sudoku puzzle grid and UNIX timestamp has to be provided as they act as the public and private keys. 
\end{enumerate}

The above two requirements must be intact such that even if the contents of the above elements are slightly changed, the whole decryption process would end up in failure, thus making the audio more distorted.

\pagebreak
For IPFS, the steps are mentioned below,
\begin{enumerate}
	\item First the algorithm process is implemented (in case of encryption), and then the audio file (encrypted) is made to pin to the IPFS nodes via the Pinata Java API. After some time, the audio file is made available globally to the IPFS network with a random IPFS hash assigned
	\item During decryption, the IPFS hash is supplied by the user in which the audio file (encrypted) will be downloaded for performing the decryption process
\end{enumerate}

\begin{figure}[htbp!]
	\centerline{\includegraphics[scale=0.5]{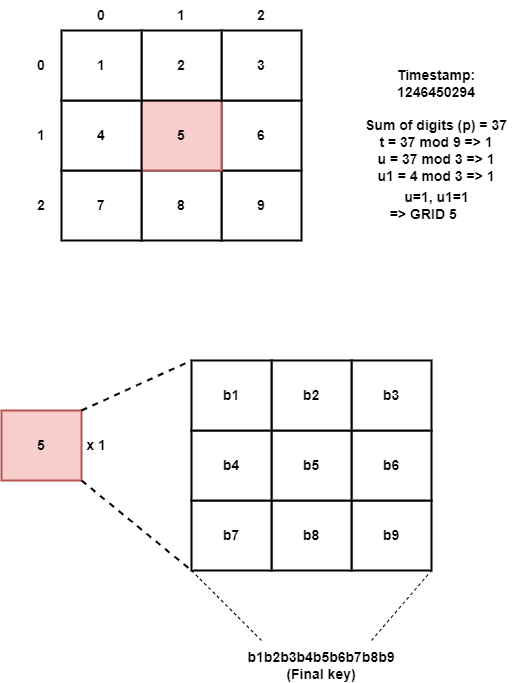}}
	\caption{Graphical representation of key generation}
	\label{diag}
\end{figure}

The overall algorithms designed for the cryptographic system are given in Alg. \ref{alg-1} and Alg. \ref{alg-2} and to understand the key generation, a sample representation has been provided in Fig. \ref{diag}

\afterpage {
\begin{algorithm}[htbp!]
	\caption{HexE - Audio Encryption/Decryption without IPFS}
	\label{alg-1}
	\begin{algorithmic}[1]
		\REQUIRE $wavFile$
        \ENSURE $wavModified$
		\STATE $Start$
		\STATE $puzzleSize \gets$ input()
		\STATE $wavAudio \gets$ input()
        \STATE $func$ generate(k)
        \{
        \IF {k equals 9}
        \STATE $GRIDSize$ = 9
        \ENDIF
        \IF {k equals 16}
        \STATE $GRIDSize$ = 16
        \ENDIF
        \IF {k equals 25}
        \STATE $GRIDSize$ = 25
        \ENDIF
        \STATE $puzzle$ = generatePuzzle(GRIDSize)
        return $puzzle$ \\
        \}
		\IF {$option \gets ENCRYPT$}
		\STATE $puzzle, timestamp \gets$ generate(k)
		\ELSE
		\STATE $puzzle, timestamp \gets$ input()
		\ENDIF
		\STATE $p \gets sumOfDigits(timestamp)$
		\STATE $t \gets$ p mod N
		\STATE $k \gets sqrt(sizeOfPuzzle)$
		\STATE $u \gets$ p mod k
		\STATE $u1 \gets lastDigit$  mod k
		\STATE $miniGrid \gets gridSelect(puzzle, u, u1)$
		\STATE $miniGrid = miniGrid * t$
		\STATE $binKey \gets$ binaryConvert($miniGrid$)
		\IF {$option \gets ENCRYPT$}
		\STATE $binArray \gets$ convertToByte($wavAudio$)
		\WHILE {$binArray.length \gets$ !end}
		\STATE $binArray[i] \gets binArray[i]$ XOR ($binkey[i]$ )
		\ENDWHILE
		\ELSE
		\STATE $binArray \gets$ convertToByte($wavAudio$)
		\WHILE {$binArray.length \gets$ !end}
		\STATE $binArray[i] \gets binArray[i]$ XOR ($binkey[i]$)
		\ENDWHILE
		\ENDIF
		\STATE $wavAudioModified \gets convertToWAV(binArray)$
		\STATE $End$
	\end{algorithmic}
\end{algorithm}
\thispagestyle{empty}
\clearpage
\begin{algorithm}[htbp!]
	\caption{HexE - Audio Encryption/Decryption with IPFS}
	\label{alg-2}
	\begin{algorithmic}[1]
		\REQUIRE $wavFile, IPFSHash$
        \ENSURE $wavModified$
		\STATE $Start$
		\STATE $puzzleSize \gets$ input()
		\STATE $wavAudio \gets$ input()
        \STATE $func$ IPFSStore(wavFile)
        \{
        \STATE $gatewayIPFS \gets$ (IPFS Gateway URL)
        \STATE $status \gets$ store(gatewayIPFS, wavFile, APIKey)
        \IF {$status$ == true}\\
        return true
        \ELSE\\
        return false
        \ENDIF
        \\
        \}
		\IF {$option \gets ENCRYPT$}
		\STATE $puzzle, timestamp \gets$ generate()
		\ELSE
		\STATE $puzzle, timestamp \gets$ input()
		\STATE $wavAudio \gets IPFSReceive(IPFSHash)$
		\ENDIF
		\STATE $p \gets sumOfDigits(timestamp)$
		\STATE $t \gets$ p mod N
		\STATE $k \gets sqrt(sizeOfPuzzle)$
		\STATE $u \gets$ p mod k
		\STATE $u1 \gets lastDigit$  mod k
		\STATE $miniGrid \gets gridSelect(puzzle, u, u1)$
		\STATE $miniGrid = miniGrid * t$
		\STATE $binKey \gets$ binaryConvert($miniGrid$)
		\IF {$option \gets ENCRYPT$}
		\STATE $binArray \gets$ convertToByte($wavAudio$)
		\WHILE {$binArray.length \gets$ !end}
		\STATE $binArray[i] \gets binArray[i]$ XOR ($binkey[i] * 5000$ )
		\ENDWHILE
		\ELSE
		\STATE $binArray \gets$ convertToByte($wavAudio$)
		\WHILE {$binArray.length \gets$ !end}
		\STATE $binArray[i] \gets binArray[i]$ XOR ($binkey[i] * 5000$ )
		\ENDWHILE
		\ENDIF
		\STATE $wavAudioModified \gets convertToWAV(binArray)$
		\IF {$option \gets ENCRYPT$}
		\STATE $HashFile \gets$ $IPFSStore(wavAudioModified)$
		\ENDIF
		\STATE $End$
	\end{algorithmic}
\end{algorithm}
\thispagestyle{empty}
\clearpage
}

\pagebreak
\section{Results}
The proposed algorithm is tested under various test cases using JUnit 5 (JUnit Jupiter) library. The algorithm handled well and the encryption, decryption process took only a few amounts of time depending on varying file size. In the test cases, security levels of 9 (9x9) and 16 (16x16) are tested. Due to current constraint limits, level 25 (25x25) is not tested. Few of the metrics are also listed down as follows.

\subsection{Encryption}
The proposed algorithm took only a few milliseconds to seconds to encrypt the contents of the WAV file depending on its file size. The audio analysis is also done in terms of SNR (Signal-To-Noise) ratio and LLR (Likelihood) ratio. To understand this, the results are plotted on Table \ref{compare1}.

\begin{table}[H]
	\caption{Time taken for encryption}
	\centering
	\begin{tblr}{|Q[c,2.5cm]|Q[c,2.5cm]|Q[c,2.5cm]|Q[c,2.5cm]|}
		\hline
		\cline{1-3} 
		\textbf{\textit{File Name}}& \textbf{\textit{File Size}}& \textbf{\textit{Security Level}} & \textbf{\textit{Time taken (ms/s)}}\\
		\hline
		Test.wav & 2 KB & 9/16 & 35ms/41ms \\
		\hline
		Test1.wav & 1 KB & 9/16 & 40ms/47ms \\
		\hline
		File1.wav & 24.5 MB & 9/16 & 715ms/1.2s \\
		\hline
		File2.wav & 35.2 MB & 9/16 & 1s/1.3s \\
		\hline
	\end{tblr}
	\label{compare1}
\end{table}

\subsection{Decryption}
The proposed algorithm took a considerable amount of time to get the byte data corrected and decrypt it to make the WAV file audible. This process also depends on the performance of the system used, so that factor is also considered here. The audio analysis is also done in terms of SNR (Signal-To-Noise) ratio and LLR (Likelihood) ratio. The performance of the decryption process is mentioned in table \ref{compare2} and for audio files, the results are tabulated in table \ref{compare3}.

\begin{table}[H]
	\caption{Time taken for decryption}
	\centering
	\begin{tblr}{|Q[c,2.5cm]|Q[c,2.5cm]|Q[c,2.5cm]|Q[c,2.5cm]|}
		\hline
		\cline{1-3} 
		\textbf{\textit{File Name}}& \textbf{\textit{File Size}}& \textbf{\textit{Security Level}} & \textbf{\textit{Time taken (ms/s)}}\\
		\hline
		Test.wav & 2 KB & 9/16 & 70ms/75ms \\
		\hline
		Test1.wav & 1 KB & 9/16 & 60ms/66ms \\
		\hline
		File1.wav & 24.5 MB & 9/16 & 1.7s/2s \\
		\hline
		File2.wav & 35.2 MB & 9/16 & 2.6s/2.8s \\
		\hline
	\end{tblr}
	\label{compare2}
\end{table}

\begin{table}[H]
	\caption{Performance analysis of files}
	\centering
	\begin{tblr}{|Q[c,2.3cm]|Q[c,2.5cm]|Q[c,2.5cm]|Q[c,2.5cm]|Q[c,2.5cm]|}
		\hline
		\cline{1-3} 
		\textbf{\textit{File Name}}& \textbf{\textit{SNR (Encrypted)}}& \textbf{\textit{LLR (Encrypted)}} & \textbf{\textit{SNR (Decrypted)}} & \textbf{\textit{LLR (Decrypted)}}\\ 
		\hline
		Test.wav & -12.0808 & -1.7098e-06 & Inf & 0 \\
		\hline
		Test1.wav & -12.0132 & -1.9750e-06 & Inf & 0 \\
		\hline
		File1.wav & -11.6495 & 4.4550e-06 & Inf & 0 \\
		\hline
		File2.wav & -11.4232 & 3.3245e-06 & Inf & 0 \\
		\hline
	\end{tblr}
	\label{compare3}
\end{table}

\begin{figure}[htbp!]
	\centerline{\includegraphics[scale=0.4]{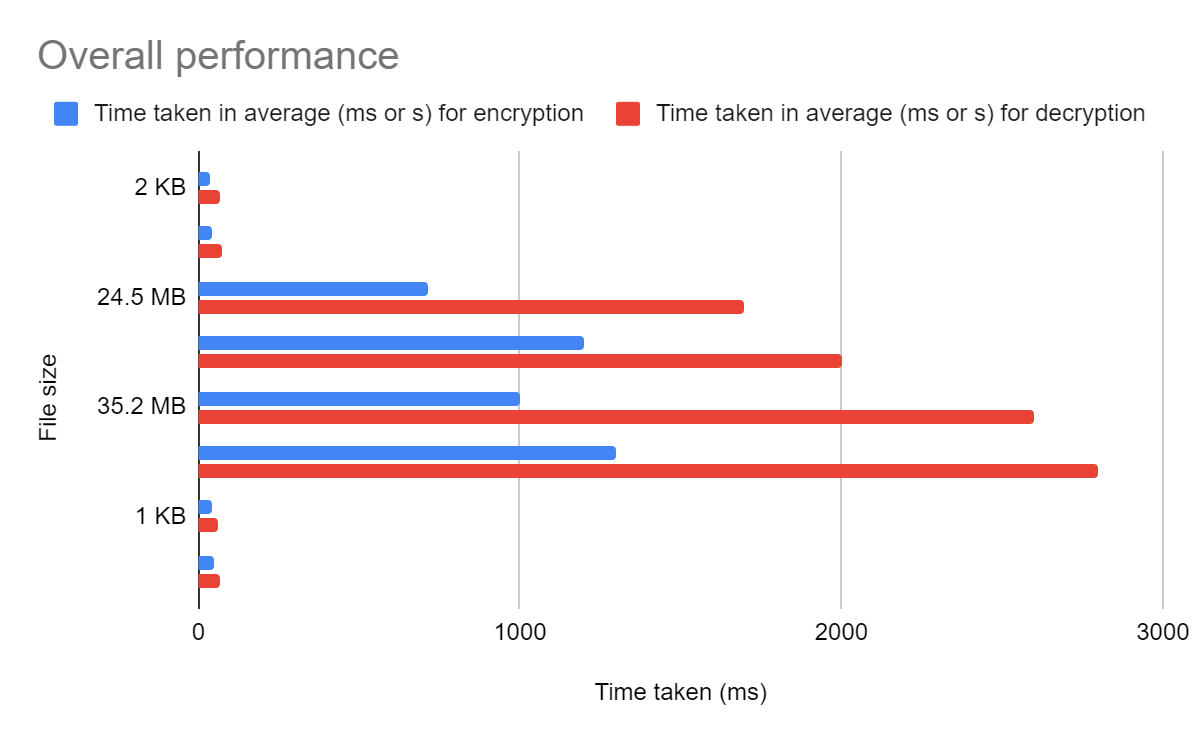}}
	\caption{Overall performance comparison of proposed algorithm}
	\label{graph}
\end{figure}

\begin{figure}[htbp!]
	\centerline{\includegraphics[scale=0.62]{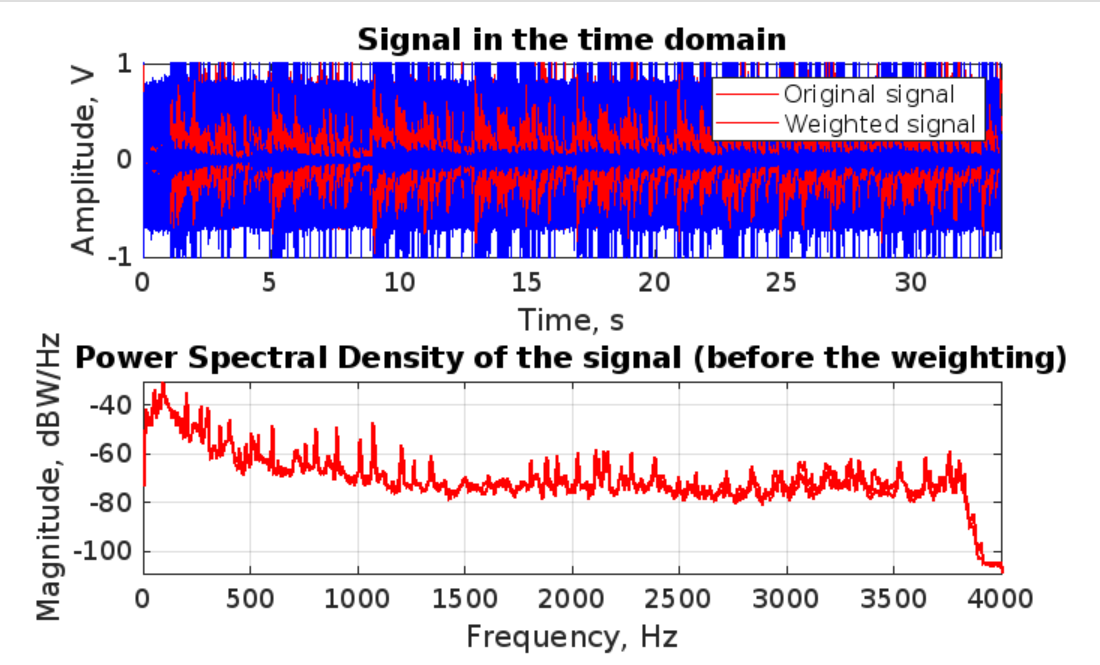}}
	\caption{Spectral analysis of original WAV file}
	\label{a1}
\end{figure}

\begin{figure}[htbp!]
	\centerline{\includegraphics[scale=0.62]{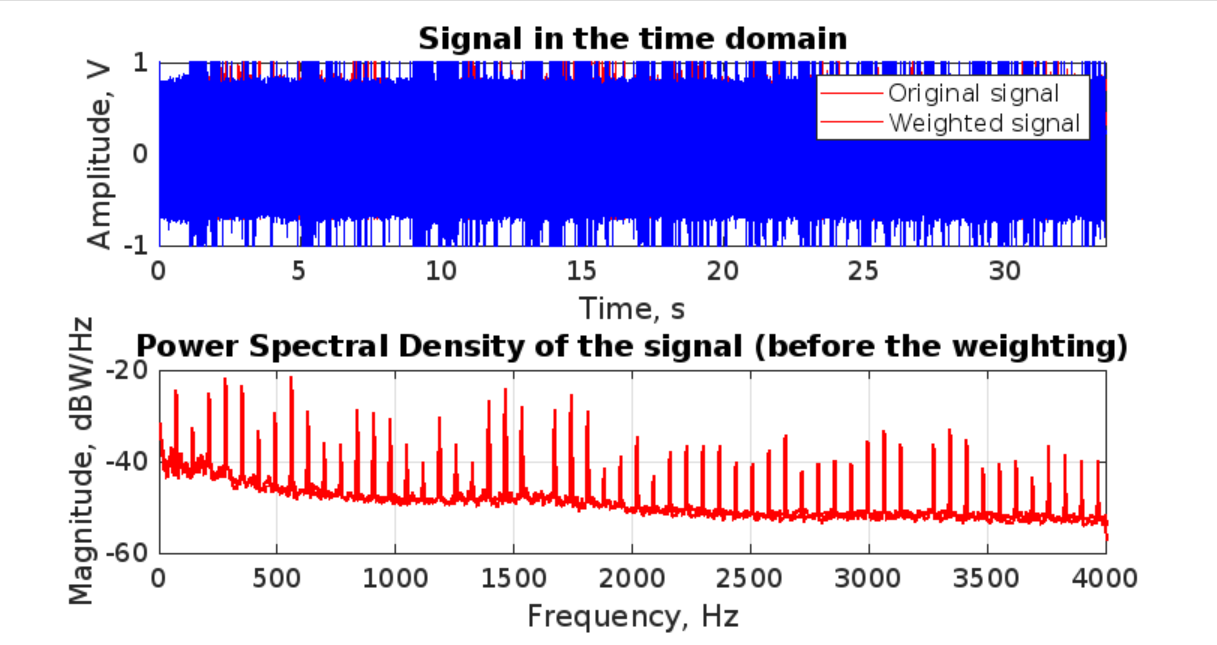}}
	\caption{Spectral analysis of encrypted WAV file}
	\label{a2}
\end{figure}

The overall performance of the proposed algorithm is analyzed and it’s found that the time taken to encrypt or decrypt an audio WAV file depends on the file content size and sample rate. The higher the file size, the higher the time taken to encrypt/decrypt the WAV file. The graph for the overall performance of the algorithm is plotted and shown in Fig. \ref{graph} and compared with existing algorithms in Table \ref{compare}. The spectral analysis of audio files is also shown in Fig. \ref{a1} and Fig. \ref{a2}

\begin{figure}[htbp!]
	\centerline{\includegraphics[scale=0.62]{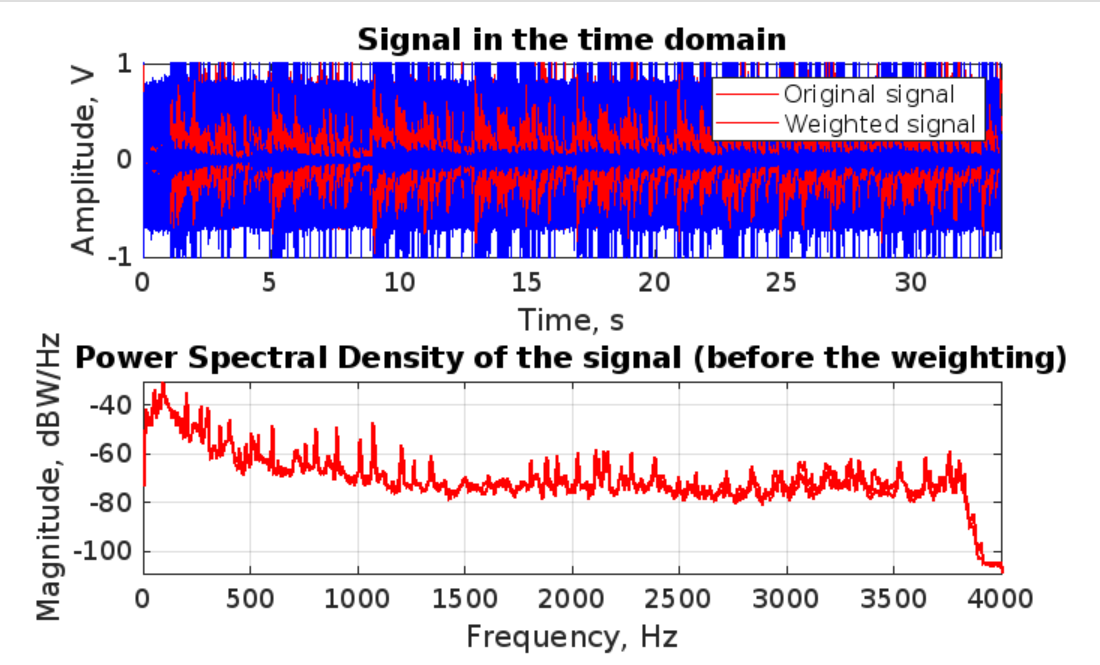}}
	\caption{Spectral analysis of decrypted WAV file}
	\label{a3}
\end{figure}

\begin{table}[H]
	\caption{Comparision with other existing algoritms}
	\centering
	\begin{tblr}{|Q[c,2cm]|Q[c,2.5cm]|Q[c,2.5cm]|Q[c,2.5cm]|Q[c,2cm]|}
		\hline
		\cline{1-3} 
		\textbf{\textit{Method}}& \textbf{\textit{Algorithm}}& \textbf{\textit{Overall time for encryption}} & \textbf{\textit{Overall time for decryption}} & \textbf{\textit{Loss in Quality}}\\
		\hline
		Proposed & NxN Sudoku puzzle with timestamp (XOR based) & 547.25 ms & 1171.375 ms & N/A \\
		\hline
		\cite{b3} & Improved Syndrome-Trellis Codes & 64 s (to embed a message) & N/A & 25\% \\
		\hline
		\cite{b6} & Modified LSB & 1438.1 ms (to insert content) & 48.8 ms (to extract content) & N/A \\
		\hline
		\cite{b10} & SuDoKu, Fuzzy C-Means and Threefish cipher & 731 ms & 1700 ms & 10\% \\
		\hline
	\end{tblr}
	\label{compare}
\end{table}
\pagebreak
\section{Conclusion}
Due to the digitalization of data, security becomes an important factor in storing any kind of data. Security is needed in audio for a variety of reasons. Audio files and streams can contain sensitive information, such as personal conversations, confidential business discussions, or proprietary information. If this audio is not properly secured, it can be intercepted and accessed by unauthorized parties, potentially leading to the disclosure of sensitive information or the misuse of confidential data. By encrypting audio and ensuring that it can only be accessed by authorized individuals, we can protect the confidentiality and integrity of the audio and the information it contains. The proposed algorithm works in two approaches: Without IPFS and With IPFS. In either of the approaches, the algorithm requires a NxN Sudoku puzzle and UNIX timestamp of the session in order for the process to execute for better results. The algorithm works in such a way that even when one bit of key is changed, the file gets distorted making it unbearable to the users. There are several potential directions for future work on this project. One potential avenue for research is to further optimize the algorithm for better performance, such as reducing the time and resource requirements for encryption and decryption. Another possibility is to explore the use of different types of puzzles or other methods of generating unique keys for encryption. Additionally, it would be interesting to examine the feasibility of implementing this approach in a real-world scenario, such as in a secure communication system. Overall, the use of sudoku puzzles and timestamps for audio cryptography has the potential to provide a reliable and secure method for transmitting sensitive audio information.

\pagebreak

\end{document}